\documentclass[11pt,a4paper,twocolumn]{article} 
\pdfoutput=1 
\usepackage{authblk} 
\usepackage{hyperref} 
\usepackage{datetime}

\title{Software must be recognised as an important output of scholarly research}
\author[1]{Caroline Jay}
\author[1]{Robert Haines}
\author[2]{Daniel S. Katz}
\affil[1]{University of Manchester, UK}
\affil[2]{University of Illinois at Urbana-Champaign, USA}
\date{\today}

\begin{document}
\maketitle

\begin{abstract}
Software now lies at the heart of scholarly research. Here we argue that as well as being important from a methodological perspective, software should, in many instances, be recognised as an output of research, equivalent to an academic paper. The article discusses the different roles that software may play in research and highlights the relationship between software and research sustainability and reproducibility. It describes the challenges associated with the processes of citing and reviewing software, which differ from those used for papers. We conclude that whilst software outputs do not necessarily fit comfortably within the current publication model, there is a great deal of positive work underway that is likely to make an impact in addressing this.
\end{abstract}

\section{Introduction}

Software is transforming scholarly research\footnote{We use ``scholarly research'' as a general term for research in science, engineering, humanities, etc.} practice, increasing the scale of knowledge production~\cite{uk_survey}, and---through the automation of analysis pipelines---putting genuine reproducibility of experiments within reach. Where once studies were conducted \emph{in vivo}, or \emph{in vitro}, they are increasingly being conducted \emph{in silico}. Software has also led to the creation of new forms of analysis and representation, enabling research or thinking that was not previously possible: computational models now form the backbone of many research domains, shifting the way in which we represent and understand the world.

Alongside the opportunities offered by computation, there is a conundrum for the research community: whilst software is now central to the production of research, it is difficult---arguably impossible---to represent it adequately in standard scholarly publications. Documents, in particular peer-reviewed papers, are currently the primary currency of scholarly research. Articles, alongside lab notes, books and reports, combine mathematical or logical formalisms with a descriptive narrative, allowing others to understand what has been discovered, and the context in which this has been achieved.

Software exists to perform processes and calculations that would otherwise be impossible in practical terms. Whilst we can endeavour to express an algorithm in pseudocode (a process fraught with problems, as the proliferation of inaccurate versions of Porter's stemming algorithm demonstrates~\cite{thimbleby2003}), many computational analyses simply cannot be translated into words or equations~\cite{jay2020challenges}. Explaining what a piece of software does will remain an essential part of reporting research, but providing access to the code itself is vital to ensuring the integrity, transparency and reproducibility of the research. This is part of the process of making the software FAIR, increasingly recognized as a key element in enabling better and more productive scholarship~\cite{Lamprecht_2020, EU-FAIR, FAIR4RSWG}.

If a computational model or analysis has complexity that cannot be adequately expressed in the form of a traditional, text-format publication, then it follows that the software should be treated as a research output in its own right, and its creators should be credited with making a contribution to scholarship. Whilst this may be acceptable in theory, the paper still rules as the primary measure of academic achievement in practice, so a rethink of how we understand and value scholarly endeavour is required. Here, we examine the reasons that software should be considered as a first class citizen of scholarly research, and outline the challenges that we must overcome to achieve this.

\section{The role of software in scholarly research}

Software is changing the way we conduct scholarly research, in terms of the sophistication of the analyses we perform and the volume of data we can process. It supports real documentation of the research process (known as provenance), and makes it possible to verify results, by improving the reproducibility of the analysis pipeline. Executable notebooks, such as Jupyter, or R markdown, are a good example of this: by interleaving explanations with software, they make it straightforward to understand and rerun the way an author has processed data. Making software methods, models and analyses open to others can greatly accelerate the rate at which we gather knowledge and make discoveries. In spite of its value, however, a great deal of research software remains unpublished and unavailable~\cite{Peng1226}. This is potentially a huge loss to scholarly research: whilst very few recent papers would exist without the aid of software, software stands on its own, and may have uses that extend far beyond a single publication~\cite{Souza2014, Souza2019}.

Currently many researchers are not working as openly as they could. The main reasons researchers give for this are embarrassment due to perceived poor quality code, a lack of confidence the software is robust for other users and usages, and the time required to prepare it for release, including the provision of appropriate licensing and documentation~\cite{jaynot}. The second point is particularly troubling: if a researcher is not confident in their own software, how can they be confident in the results it produces? Improving the visibility, and therefore scrutiny, of research software would mitigate these problems, increasing both the openness of a project, and the confidence in its conclusions. It is important to note here that valuing the software in its own right is an important catalyst to good development. Where the software is simply regarded as a means to an end, rather than an integral part of the research, the temptation to minimize the time and resources that go into its creation is high.

Increased openness may be viewed by some as a burden, but it ultimately has the potential to benefit researchers and the culture they work in. A report from the UK Parliamentary Office of Science and Technology, ``Integrity in Research''~\cite{post544}, puts an emphasis on enforcing the integrity of research outcomes, potentially via regulation, but does not address \emph{how} researchers' everyday practices should evolve to ensure this outcome is achieved. Telling researchers that they are not working with integrity---in effect that they are not doing research \emph{well}---is applying pressure in the wrong place; while mistakes happen, the vast majority of researchers are working honestly. Instead, a focus on promoting openness is likely to have a much larger impact while fixing the same problem, as it will naturally increase the chances of mistakes being caught. Valuing software in its own right, and giving credit to those who produce it, is an excellent way of motivating this shift in practice.

\section{When is software an output?}

Software plays different roles in the research process. It can be a tool for supporting the work---software as infrastructure---or embody the research itself---for example, in a scientific simulation.  The role of the same piece of software can vary according to the context. To a computer scientist in the field of workflow management, the workflow software would be considered a direct output, as it is the manifestation of the research. To a biologist, this same software would be considered a tool: useful for analyzing results, but not itself an output of the research. For a bioinformatician, both using and developing the tool, the answer is somewhere in the middle: whilst the core research may be in the life science domain, the modifications made to the tool as a result of this work could also be considered an output, advancing workflow management~\cite{jay2016}.

Drawing a hard line between these categories is difficult. Another way of considering software within the research process is from the angle of reproducibility and reusability. If any bespoke software is developed as part of the research, even if it is just an analysis script, then making it available is an important part of the reproducibility pipeline. This is only part of the challenge however; to maintain the integrity of the software as a part of the research process, it is important not just to be able to access it, but also to be able to refer to it accurately.

\section{Citing software}

Many venues now mandate that data, and increasingly analysis software, be archived and made available alongside a paper~\cite{katz2020}, in a welcome step towards improving the reproducibility of research. This works well when the software is a straightforward analysis script, but the process of archiving quickly becomes complex with anything beyond this. A preserved `snapshot' of the environment in which a discovery was made is crucial to fully understanding the provenance and reliability of the data, and the potential permanence of software promises to greatly increase the rigour of the scientific process. Most publication venues lack guidelines that encourage citing software directly, however, and doing so is not general practice. A common workaround is to cite a related paper instead. This might be a paper describing a larger study, where the software was integral to that research and is described in the methods section, or it might be a ``software paper'': a paper that exists solely to describe the software, in a venue such as SoftwareX, the Journal of Open Research Software (JORS), the Journal of Open Source Software (JOSS) or F1000 Research. In either case the software referred to in those papers will be out of date very quickly. Software does not stay still---bugs are fixed, new functionality is added and optimizations are made---and development is rarely paused for lengthy journal submission processes to complete. The specific release of software must be preserved (archived) and then cited directly, in each publication in which it is used, to be sure that the correct version is referenced each time, and can be used for reproducibility. Providing information that will help people find the latest version of the software in a repository is also helpful, as this may be the one most useful to someone who wishes to use or develop the software further~\cite{Katz-repositories-blog}.

Precisely how to cite archived software remains an open question~\cite{smith2016}, but an obvious mechanism for doing this is to use a Digital Object Identifier (DOI) for the particular version of the software, and include this in the reference list in the paper. As software and papers have a symbiotic relationship, it would be ideal to link back to the paper from the software. The publication workflow makes this difficult, however, as the paper will be published after the software, and at that point it is not possible to alter the software object and maintain the integrity of the DOI. Indeed, the nature of the DOI allocation process means that it is impossible for two objects to reference each other without careful planning and DOI reservation. This demonstrates the necessity for software to be considered an object in its own right, standing alone and independently of any paper.

\section{Peer review of software}

If software is to be considered an output of scholarly research it is important to ensure, as with text-based publications, it is valid and reliable. Peer review is currently the accepted method for determining the validity (and to some extent, value) of research outputs, and the format for the review of text publications is well-established. `Software-paper' venues have a review process for software, but the methodology currently followed often focuses primarily on checking that the software meets technical requirements (for example, that it is open source and has installation instructions), rather than fully evaluating its scientific contribution. Clear documentation, strategies for quality assurance, such as unit tests, and following relevant coding standards are indicators of rigour, but should be treated as proxies, rather than guarantees of this.

Code review---checking that the way in which software is written meets certain quality standards---is widely used in industry to check for defects, and ensure that software is efficient and usable by others. This process, analogous to checking that a paper is free from language errors, and that the narrative is unambiguous, has an important role in assessing research software, where accuracy is of paramount importance. Code review is an extremely time consuming process, however, particularly where the reviewer is unfamiliar with the software, and as such realising this will be a challenge. Determining the scholarly `contribution' of software as a research output (which remains a contentious issue for traditional publications) may be less important if we take the view that its value can be judged by the papers in which it is cited, or the number of people who go on to use or extend it.

\section{Conclusion}

Software is now integral to scholarly research, and it is thus essential that it is open, accessible, and valued by the research community. The present publication model falls short of guaranteeing any of these things, but a shift is gradually occurring. Peer review of software is likely to remain a challenge, and may require a different approach from that used for papers. Official recognition of software as a research output will ultimately be transformative, improving the quality, reproducibility and scalability of our knowledge production, as well as recognising the often hidden role of the increasing number of scholarly researchers who spend most of their time writing code.

\bibliographystyle{IEEEtran} 
\bibliography{references}

\end{document}